\documentclass[12pt]{article}

\usepackage{amssymb,amsmath}
\usepackage{graphicx}
\usepackage{subfigure}

\usepackage{cite}

\newcommand{\be}{\begin{equation}}
\newcommand{\ee}{\end{equation}}

\newcommand{\bt}{\beta}

\newcommand{\ep}{\varepsilon}
\newcommand{\al}{\alpha}
\newcommand{\ra}{\rightarrow}

\newcommand{\gm}{\gamma}

\begin{document}

\begin{center}

{\Large{\bf 

        Self-similar continued root approximants}  \\ [5mm]

  S. Gluzman, V.I. Yukalov$^*$ } \\ [3mm]

{\it Bogolubov Laboratory of Theoretical Physics, \\ 
Joint Institute for Nuclear Research, Dubna 141980, Russia }

\end{center}

\vskip 5cm

\begin{abstract}
A novel method of summing asymptotic series is advanced. Such series repeatedly 
arise when employing perturbation theory in powers of a small parameter 
for complicated problems of condensed matter physics, statistical physics, and 
various applied problems. The method is based on the self-similar approximation 
theory involving self-similar root approximants. The constructed self-similar 
continued roots extrapolate asymptotic series to finite values of the expansion 
parameter. The self-similar continued roots contain, as a particular case, 
continued fractions and Pad\'{e} approximants. A theorem on the convergence of 
the self-similar continued roots is proved. The method is illustrated by several 
examples from condensed-matter physics.
\end{abstract}

\vskip 3cm


\vskip 3cm

$^*$Corresponding author.

{\bf E-mail address}: yukalov@theor.jinr.ru (V.I. Yukalov)

\newpage

\section{Introduction}

The standard difficulty, repeatedly arising in various problems of condensed-matter
physics, statistical and chemical physics, field theory, and many other problems
of theoretical physics and applied mathematics, is the necessity of using perturbation
theory in powers of a parameter assumed to be asymptotically small, while in reality 
this parameter is either finite or even tending to infinity. How then it would be 
possible to extrapolate the asymptotic series of perturbation theory to the finite,
or even to infinite values of the parameter? The most often used methods of extrapolation
are based on Pad\'{e} approximants [1]. 

But, when extrapolating an asymptotic expansion $f_k(x)$ to large values of the 
parameter $x$, by means of the Pad\'{e} approximants, one meets the ambiguity, since   
the straightforward use of these approximants yields
$$
P_{M/N}(x) \sim x^{M-N} \qquad (x\ra\infty) \;   ,
$$
which, depending on the relation between $M$ and $N$, can tend to either
infinity (when $M > N$), to zero (when $M < N$), or to a constant (if $M = N$).

When the character of the large-variable limit is known, one can invoke the
two-point Pad\'{e} approximants [1]. However the accuracy of the
latter is not high and their definition contains several drawbacks. (i) First of 
all, when constructing these approximants, one often obtains spurious poles 
yielding unphysical singularities, sometimes with a large number of poles [1,2]. 
(ii) Second, there are the cases when Pad\'{e} approximants are not able to sum 
perturbation series even for small values of an expansion parameter [3].
(iii) Third, in the majority of cases, for achieving a reasonable accuracy, one 
needs to have tens of terms in perturbative expansions [1], while interesting 
problems provide, as a rule, only a few terms. (iv) Fourth, defining the two-point 
Pad\'{e} approximants, one always confronts the ambiguity in distributing the 
coefficients for deciding which of these must reproduce the left-side expansion and 
which the right-side series. This ambiguity aggravates with the increase of the 
approximants orders, making it difficult to compose two-point Pad\'{e} tables. For 
the case of a few terms, this ambiguity makes the two-point Pad\'{e} approximants 
practically unapplicable. For example, it has been shown [4] that, for the
same problem, one may construct different two-point Pad\'{e} approximants,
all having correct left and right-side limits, but differing from each other
in the intermediate region by 1000$\%$ of uncertainty. Hence, in the case of short 
series the two-point Pad\'{e} approximants do not allow for getting a reliable 
description. (v) Fifth, the two-point Pad\'{e} approximants can be used for 
interpolating between two different expansions not always, but only when these two 
expansions enjoy compatible variables [1]. When the expansions have incompatible
variables, the two-point Pad\'{e} approximants cannot be defined in principle.
(vi) Sixth, interpolating between two points, one of which is finite and another
is at infinity, one is able to characterize the large-variable limit of only
rational powers [1]. (vii) Finally, it may happen that in the large-variable 
limit only the power is known, while the amplitude is not. Then the two-point 
Pad\'{e} approximants cannot be defined. 

There exists a more general approach for extrapolating asymptotic series in powers
of a small parameter, or a variable, to finite and even infinite values of such 
variables. This approach is based on the self-similar approximation theory [5-17].
In the frame of this theory, we have developed the methods of extrapolating 
asymptotic series by using several types of self-similar approximants, such as
optimized approximants, nested exponentials, nested roots, iterated roots, and 
factor approximants [5-17].

In the present paper, we advance a novel type of self-similar approximants that
may be called {\it self-similar continued root approximants}, or, for short,
{\it self-similar continued roots}. In a particular case, these continued roots 
reduce to continued fractions [18] and, respectively, to Pad\'{e} approximants. 
But, generally, their form is different and not reduceable to continued fractions.    
The self-similar continued roots could be transformed into expressions of the type
of the numerical nested radicals [19-22], which, however, is not convenient for the 
extrapolation procedure applied to functions.

In Sec. 2, we explain how the self-similar continued roots arise in the process
of the self-similar renormalization of asymptotic series and prove the convergence 
of these root approximants. In Sec. 3, we demonstrate, by several examples from
condensed-matter physics, that the continued roots can be employed as approximants
extrapolating asymptotic series and providing good accuracy. Possible generalizations
for the continued root approximants are also mentioned.

\section{Construction and convergence of self-similar \\ continued roots}

Assume that we are looking for the solution of a complicated problem characterized
by a real function $f(x)$ of a real variable $x \in \mathbb{R}$. And let this
problem be not solvable explicitly, but allowing only for an approximate solution 
at small values of $x$, 
\be
\label{1}
 f(x) \simeq f_k(x) \qquad (x\ra 0 ) \;  ,
\ee
where it is represented by asymptotic series of orders $k = 1,2,\ldots$,
\be
\label{2}
 f_k(x) = \sum_{n=0}^k a_n x^n \qquad (a_0 = 1 ) \;  .
\ee
Here, without the loss of generality, we set $a_0 = 1$. This is because any function
$$
  g(x) \simeq g_k(x) \qquad ( x \ra 0) \; ,
$$
that at small $x$ is given by the series 
$$
 g_k(x) = g_0(x) \sum_{n=0}^k a_n x^n \;  ,
$$
can always be written in such a form, where $a_0 = 1$. Here $g_0(x)$ is assumed to 
be known. Then the considered function is defined as 
$$
f(x) \equiv \frac{g(x)}{g_0(x)} \;   .
$$
Respectively, the series at small $x$ take the form
$$
f_k(x) = \frac{g_k(x)}{g_0(x)} = \sum_{n=0}^k a_n x^n \; ,
$$
corresponding to Eq. (2) with $a_0 =1$.    

To apply the self-similar renormalization procedure [5-17] to series (2), recall 
that the first-order series, having the linear from $f_1(x) = 1 + a_1 x$, transforms 
into $f^*_1(x) = (1 + A_1 x)^s$. Notice that series (2) can be rewritten in the form
$$
 f_k(x) = \left ( 1 + a_1 x \left ( 1 + \frac{a_2}{a_1}\; x \left ( 1 + 
\frac{a_3}{a_2}\; x \left ( 1 + \ldots + \frac{a_{k-1}}{a_{k-2}} \; x \left ( 1 + 
\frac{a_k}{a_{k-1}} \; x \right ) \right ) \right ) \right ) \ldots \right ) \;  .
$$
Then applying the self-similar renormalization sequentially to each of the terms 
inside the brackets, we come to the {\it self-similar continued root} 
\be
\label{3}
 f_k^*(x ) = \left ( 1 + A_1 x \left ( 1 + A_2 x \ldots \left ( 1 + A_k x \right )^s
\right )^s \ldots \right )^s \;  .
\ee
For instance, in low orders we have
$$
f_1^*(x) = ( 1 + A_1 x)^s \; , \qquad 
f_2^*(x) = \left ( 1 + A_1 x \left ( 1 + A_2 x\right )^s \right )^s \; ,
$$
$$
 f_3^*(x) = \left ( 1 + A_1 x \left ( 1 + A_2 x \left ( 1 + A_3 x 
\right )^s \right )^s \right )^s \;  .
$$
Note that, generally speaking, instead of the same power $s$, we could take 
different powers $s_k$. This, however, would increase the number of parameters
that need to be defined, which would require to have more information on the 
sought function. The use of the same powers $s$ simplifies the problem. 

One can notice that in the particular case of $s = - 1$, the self-similar continued 
roots reduce to the continued fractions [18] and, respectively, to the Pad\'{e}
approximants. Thus,
$$
f_{2k}^*(x) = P_{k/k}(x)  \; , 
$$
$$
f_{2k+1}^*(x) = P_{k/k+1}(x) \qquad ( s = -1 ) \;    .
$$
In this way, the self-similar continued roots (3) generalize the notion of
continued fractions.   
 
It is easy to see that if we require the continued roots to be real-valued 
quantities for all $x \in [0, \infty)$, the parameters $A_n$ have to be 
non-negative,
\be
\label{4}
A_n \geq 0 \qquad ( n = 1,2, \ldots, k ) \;   .
\ee

The parameters $A_n$ are to be defined by the accuracy-through-order procedure,
by comparing the like orders in the small-variable expansion for the continued 
roots (3) with the corresponding series (2), 
\be
\label{5}
 f_k^*(x) \simeq f_k(x) \qquad ( x \ra 0 ) \;  .
\ee
For instance, in the first order, the expansion of approximant (3) is
$$
 f_1^*(x) \simeq 1 + s A_1 x \qquad ( x \ra 0 ) \;  ,
$$
which gives 
$$
A_1 = \frac{a_1}{s} \;   .
$$
In the second order, at small $x$, we have
$$
f_2^*(x) \simeq 1 + sA_1 x + \frac{s A_1}{2} \; ( s A_1 - A_1 + 2s A_2 ) x^2 \;  ,
$$
from where we get the same parameter $A_1$ and 
$$
A_2 = \frac{(1-s)a_1^2 + 2s a_2}{2s^2a_1} \; .
$$
   
What is left to be defined is the power $s$. Assume that the large-variable
asymptotic behaviour of the sought function is known to be
\be
\label{6}
 f(x) \simeq B x^\bt \qquad ( x \ra \infty) \;  .
\ee
In turn, the large-variable behaviour of approximant (3) is
\be
\label{7}
  f_k^*(x) \simeq B_k x^{\bt_k} \qquad ( x \ra \infty) \; ,
\ee
with the amplitude
\be
\label{8}
  B_k = \prod_{n=1}^k A_n^{s^n} 
\ee
and the power
\be
\label{9}
 \bt_k = \sum_{n=1}^k s^n = \frac{s-s^{k+1}}{1-s} \;  .
\ee
Let us assume (which will be proved below) that the approximants converge, 
under $k \ra \infty$, so that there exists the limit 
\be
\label{10}
 \lim_{k\ra\infty} \bt_k = \bt \;  .
\ee
This requires that the power $s$, by modulus, be smaller than one, as a 
result of which, Eqs. (9) and (10) yield
\be
\label{11}
\bt = \frac{s}{1-s} \qquad (|s| < 1 ) \;  .
\ee
Inverting the latter equality gives
\be
\label{12}
 s = \frac{\bt}{1+\bt} \qquad \left ( \bt > -\; \frac{1}{2} \right ) \; .
\ee
    
We have assumed above that the sequence of approximants (3) converges as
$k$ increases. Now, we provide the proof of this.   

\vskip 2mm

{\bf Theorem}. Suppose that $x$ lays in a finite interval
\be
\label{13}
 x\in [ 0, L] \qquad ( L < \infty ) \;  ,
\ee
the power $s$ is such that $|s| < 1$ and the parameters $A_n$ are non-negative 
and bounded,
\be
\label{14}
 0 \leq  A_n \leq M \qquad ( n = 1,2,\ldots ) \;  .
\ee
Then the sequence of approximants (3) converges,
\be
\label{15}
\lim_{k\ra\infty} f_k^*(x) = f_\infty^*(x) \;   .
\ee
 
\vskip 2mm

{\it Proof}. The self-similar continued root (3) can be reduced to the form
of the nested radicals
$$
f_k^*(x) = \left ( 1 + \left ( x_2 + \left ( x_3 + \ldots x_{k-1}^s 
\right )^s \right )^s \ldots \right )^s \;   ,
$$
in which the notation 
$$
x_n \equiv x^{\gm_n} \prod_{j=1}^{n-1} A_j^{s^{j-n} } \qquad 
(x_1 \equiv 1)
$$
is used, where
$$
 \gm_n \equiv \frac{1-s^{n-1}}{(1-s)s^{n-1}} \qquad 
( n = 2,3,\ldots ) \;  .
$$
By the Herschfeld theorem [19], the sequence of the nested radicals, with 
non-negative $x_n$, converges if and only if all $x_n^{s^n}$ are bounded. 
Condition (14) shows that
$$
  \prod_{j=1}^{n-1} A_j^{s^j }  \leq M^{\gm_n s^n} \; ,
$$
with 
$$
 \gm_n s^n = \frac{s-s^n}{1-s} \;  .
$$
Taking into account condition (13), we see that the terms $x_n^{s^n}$ are bounded 
for all finite $n$,
$$
 x_n^{s^n} \leq (LM)^{\gm_n s^n} \;  .
$$
When $n \ra \infty$, under $|s| < 1$, then 
$$
 x_n^{s^n} \leq (LM )^{s/(1-s)} \qquad ( n \ra \infty) \;  .
$$
Therefore all terms $x_n^{s^n}$ are bounded for any $n$, including $n \ra \infty$.
Hence the sequence of the nested radicals converges, as well as the sequence of
the self-similar continued roots (3).     

\vskip 3mm

{\it Remark}. When considering the large-variable behaviour, it is always possible 
to treat this variable as asymptotically large, but finite, so that condition (13)
be valid.

\section{Extrapolation by means of self-similar continued roots}

The self-similar continued roots (3) can be used for approximating functions
in the whole region of the variable, between zero and infinity. Below, we 
illustrate that these approximants extrapolate, with a good accuracy, the 
small-variable asymptotic expansion (2) to the large-variable region, where
the sought function behaves as in Eq. (6). Defining the power $s$ according 
to Eq. (12), we shall calculate the amplitude $B_k$, evaluating the accuracy 
of the found approximations (8) by comparing it with the known value $B$.

\subsection{Nonlinear Schr\"{o}dinger equation}

Many problems in quantum physics, condensed-mater physics, and optics reduce
to the nonlinear Schr\"{o}dinger equation of the type
\be
\label{16}
 -\; \frac{1}{2} \; \frac{d^2\psi}{dx^2} + \left ( \frac{1}{2} \; x^2 +
 g |\psi|^2 \right ) \psi = E\psi \;  .
\ee
Here we consider the one-dimensional case, where $x \in (-\infty, \infty)$.
The first term corresponds to kinetic-energy operator. The second term 
represents a harmonic external potential. The third term describes the
interactions of atoms, typical, e.g., of trapped atomic gases [23-25],
with a coupling parameter $g$. 

The ground-state solution represents equilibrium Bose-Einstein condensate. 
And the higher-energy solutions characterize coherent modes [25]. The 
spectrum of coherent modes for Eq. (16) can be written [13] in the form
\be
\label{17}
 E_n = \left ( n + \frac{1}{2} \right ) f(\al) \;  ,
\ee
where $n = 0, 1, 2, \ldots$ and the function $f(\alpha)$ can be found 
by means of perturbation theory as an expansion
\be
\label{18}
f_k(\al) = 1 + \sum_{m=1}^k a_m \al^m
\ee
in powers of the effective coupling parameter
\be
\label{19}
 \al \equiv \frac{2J_n}{1+2n} \; g \;  ,
\ee
in which 
$$
 J_n \equiv \frac{1}{2^n\pi n!} \; \int_{-\infty}^\infty
H_n^4(x) e^{-2x^2} \; dx \;  ,
$$
with $H_n(x)$ being Hermite polynomial. The first five coefficients in 
expansion (18) are
$$
a_1 = 1 \; , \qquad a_2 = -\; \frac{1}{8} \; , \qquad 
a_3 = \frac{1}{32} \; , \qquad
a_4 = -\; \frac{1}{128} \; , \qquad a_5 = \frac{3}{2048} \;   .
$$

At large $\alpha$, we have
\be 
\label{20}
 f(\al) \simeq \frac{3}{2} \; \al^{2/3} \qquad (\al\ra\infty) \;  .
\ee
Hence, $\beta = 2/3$, which, in view of Eq. (12), gives $s = 2/5$.

The continued roots (3), at large $\alpha$, lead to the form
\be
\label{21}
f_k^*(\al) \simeq B_k \al^{2/3} \;   .
\ee
For the amplitude we find the approximations
$$
B_2 = 1.549484 \qquad (3.3\%) \; ,
$$
$$
 B_3 = 1.554034 \qquad (3.6\%) \;  ,
$$
$$
B_4 = 1.539048 \qquad (2.6\%) \; ,
$$
$$
 B_5 = 1.523475 \qquad (1.6\%) \;  ,
$$
where in brackets the related percentage errors are shown.

\subsection{Fr\"{o}hlich optical polaron}

The ground-state energy of the Fr\"{o}hlich optical polaron can be written
as a function
\be
\label{22}
E(\al) = -\al f(\al)
\ee
of the effective coupling parameter $\alpha$. The function $f(\alpha)$ here
can be calculated in the second-order perturbation theory giving [26] the
expression
\be
\label{23}
f_2(\al) = 1 + a_1 \al + a_2 \al^2 \;   ,
\ee
with the coefficients 
$$
 a_1 = 1.591962\times 10^{-2} \; , \qquad
 a_2 = 0.806070 \times 10^{-3} \;   .
$$
The strong-coupling asymptotic behavior was found by Miyake [27,28] as
\be
\label{24}
  f(\al) \simeq B\al \qquad (\al \ra\infty) \; ,
\ee
with the amplitude $B = 0.108513$. This gives for the ground-state energy (22)
\be
\label{25}
 E(\al) \simeq - B\al^2 \qquad (\al \ra\infty ) \; .
\ee

For function (24), we have $\beta = 1$, hence $s = 0.5$. Extrapolating expansion
(23) by the second-order continued root, we get $B_2 = 0.1044$, with the error 
of $3.8 \%$.

\subsection{Fluctuating fluid membrane}

The pressure of a fluctuating fluid membrane, as a function of stiffness $g$ can
be represented in the form
\be
\label{26}
 P(g) = \frac{\pi^2}{8g^2}\; f(g) \;  .
\ee
Perturbation theory for the function $f(g)$ has been done [29] up to the six-th 
order, resulting in the expansion
\be
\label{27}
 f_k(g) = 1 + \sum_{n=1}^k a_n g^n \;  ,
\ee
with the coefficients
$$
a_1 = \frac{1}{4} \; , \qquad a_2 = \frac{1}{32} \; , \qquad
a_3 = 2.176347 \times 10^{-3} \; ,
$$
$$
a_4 = 0.552721 \times 10^{-4} \; ,  \qquad 
a_5 = -0.721482 \times 10^{-5} \; ,  \qquad 
a_6 = -1.777848 \times 10^{-6} \; .
$$
  
The pressure of the membrane between hard walls corresponds to the stiffness 
$g \ra \infty$. The hard-wall limit was calculated by Monte Carlo techniques 
[30], yielding   
\be
\label{28}
 P(\infty) = 0.0798 \pm 0.0003 \;  .
\ee
This corresponds to the asymptotic behaviour of the function $f(g)$ as 
\be
\label{29}
 f(g) \simeq 0.064683 g^2 \qquad (g \ra \infty ) \;  .
\ee
Therefore here $\beta = 2$ and $s = 2/3$.  

Employing the self-similar continued roots, we find the following approximations 
for the hard-wall pressure:
$$
P_2^*(\infty ) = 0.047705 \qquad (-40\% ) \; ,
$$
$$
P_3^*(\infty ) = 0.061904 \qquad (-22\% ) \; ,
$$
$$
P_4^*(\infty ) = 0.072407 \qquad (-9.3\% ) \; ,
$$
$$
P_5^*(\infty ) = 0.079569 \qquad (-0.29\% ) \; ,
$$
$$
 P_6^*(\infty ) = 0.083702 \qquad (4.9\% ) \;  ,
$$
where in brackets the corresponding percentage errors are given.

\subsection{Fluctuating fluid string}

The free energy of a fluctuating fluid string coincides with the ground-state 
energy of a particle in a box [31,32]. The latter, as a function of the wall 
stiffness $g$, can be written as 
\be
\label{30}
 E(g) = \frac{\pi^2}{8g^2} \; f(g) \;  ,
\ee
with the notation 
\be
\label{31}
f(g) = 1 + \frac{g^2}{32} + \frac{g}{4} \; \sqrt{1 + \frac{g^2}{64} } \; .
\ee

Perturbative expansion in powers of $g$ yields
\be
\label{32}
f_k(g) = 1 + \sum_{n=1}^k a_n g^n \;   ,
\ee
with the coefficients
$$
a_1 = \frac{1}{4} \; , \qquad a_2 = \frac{1}{32} \; , \qquad 
a_3 = \frac{1}{512} \; , \qquad a_4 = 0 \; ,
$$
$$
 a_5 = -\; \frac{1}{131072} \; , \qquad a_6 = 0 \; , \qquad 
a_7 = \frac{1}{16777216} \;  ,
$$
and so on.  

The case of rigid walls corresponds to $g \ra \infty$, which results in
\be
\label{33}
 E(\infty) = \frac{\pi^2}{128} = 0.077106 \;  .
\ee
This implies the asymptotic behaviour of function (31) as
\be
\label{34}
 f(g) \simeq 0.0624998 g^2 \qquad (g\ra \infty) \;  .
\ee
From here, we have $\beta = 2$ and $s = 2/3$. 

We calculate the rigid-wall limit by extrapolating expansions (32) by means
of the self-similar continued roots. The results for different approximations
$E_k^*(\infty)$  are presented in the Table, together with the related 
percentage errors
\be
\label{35}
 \ep_k \equiv \frac{E_k^*(\infty) - E(\infty)}{E(\infty)} \;
\times 100\% \;  .
\ee
 
\vskip 1cm

\begin{tabular}{|l|l|r|} \hline
$ k $& $ E^*_n(\infty)$ & $\ep_k$  \\ \hline
2    &  0.047705    &  $-$38\%      \\ \hline
3    &  0.061362    &  $-$20\%      \\ \hline
4    &  0.070598    &  $-$8.4\%      \\ \hline
5    &  0.076155    &  $-$1.2\%      \\ \hline
6    &  0.079097    &  2.6\%         \\ \hline
7    &  0.080336    &  4.2\%         \\ \hline
8    &  0.080533    &  4.4\%         \\ \hline
9    &  0.080141    &  3.9\%         \\ \hline
10   &  0.079540    &  3.2\%         \\ \hline
11   &  0.079199    &  2.7\%         \\ \hline
12   &  0.079123    &  2.6\%         \\ \hline
13   &  0.078363    &  1.6\%     \\ \hline
\end{tabular}

\vskip 1cm
{\parindent=0pt
{\bf Table}: Self-similar continued root approximants of different orders $k$ 
for the hard-wall energy $E_k^*(\infty)$, with the related percentage errors. }

\subsection{Discussion and possible generalization}

The problem is considered of extrapolating asymptotic series in powers of a 
small variable to large values of this variable.  A new structure is shown to 
arise as a result of the self-similar renormalization, giving the self-similar
continued root approximants. The use of these approximants for extrapolation 
procedure requires the knowledge of the power for the large-variable behaviour.  
The convergence of the sequence of these approximants is proved. Several 
examples from quantum condensed matter physics illustrate good accuracy of the 
approximations. These examples show that convergence is not necessarily monotonic, 
but can be oscillating. Nevertheless, the convergence theorem tells us that 
the sequence of these approximants does converge.  

The method is applicable, when all parameters $A_n$ in form (3) are non-negative.
It may happen that for particular cases some of these parameters occur to become 
negative, which would yield to complex expressions for the roots. In that 
situation, the method cannot be used directly, but one can proceed by taking 
the highest available real-valued root approximation and using the following 
terms for constructing corrected approximants as is explained in Ref. [17].

We have checked the applicability of the method for a number of other problems.
As a rule, the low-order approximants are usually real, giving good accuracy, 
but may become complex for higher-order approximants. For instance, if we try
to approximate, by continued roots, the statistical sum of the zero-dimensional 
oscillator as a function of a coupling parameter [33], when $\beta = -1/4$ 
and $s = - 1/3$, we get in the second order the strong-coupling amplitude 0.970, 
whose error is $5\%$. The third and fourth approximants are real, though slightly 
worse than the second one. However, the fifth approximant is complex. 

In another example, we calculate the expansion factor of a three-dimensional 
polymer chain as a function of an effective coupling parameter [34,35], when 
$\beta = 0.3544$ and $s = 0.261666$. The second-order approximant for the 
amplitude of the strong-coupling behavior gives 1.554, which is within an error 
of $1.5\%$. The third-order approximant is yet real, but the fourth is complex. 
Thus, in the above two examples, we need to limit calculations by the low-order 
approximants. However, it is worth recalling that the majority of the most 
interesting problems in condensed-matter physics rarely enjoy the luxury of 
having many expansion terms. The standard situation is when one is able to 
derive just a few lowest terms of perturbation theory.     

In the main expression (3) that has been used throughout the paper, we have taken
the same powers $s$. Strictly speaking, it has been admissible to write a more
general expression 
$$
f_k^*(x) = \left ( 1 + A_1 x \left ( 1 + A_2 x \ldots \left ( 1 + A_k x 
\right )^{s_k}  \right )^{s_{k-1}} \ldots \right )^{s_1} \;   ,
$$
with different powers $s_n$. But then, it would be necessary to have more 
information on the large-variable behaviour in order to define all these different
powers. 

Concluding, we have suggested a novel method for extrapolating the small-variable
asymptotic series to the large values of the variable. The method produces a 
convergent sequence of approximants and can be used for extending the validity
of perturbation theory.  

\vskip 5mm

{\bf Acknowledgment}

\vskip 3mm
One of the authors (V.I.Y.) is grateful for many useful discussions and help to 
E.P. Yukalova. Financial support from the Russian Foundation for Basic Research 
is appreciated.

\end{document}